\documentstyle[aaspp4]{article} 

\newcommand\cE{{\cal E}}
\newcommand\cL{{\cal L}}

\newcommand\simlt{\lower.5ex\hbox{$\; \buildrel < \over \sim \;$}}
\newcommand\simgt{\lower.5ex\hbox{$\; \buildrel > \over \sim \;$}}

\begin{document}

\title{General Relativistic, Neutrino-Assisted MHD winds
- Theory and Application to GRBs. I. Schwarzschild Geometry }
\author{Amir Levinson\altaffilmark{1}}
\altaffiltext{1}{School of Physics \& Astronomy, Tel Aviv University,
Tel Aviv 69978, Israel; Levinson@wise.tau.ac.il}
\begin{abstract} 
A model for GRMHD disk outflows with neutrino-driven mass ejection is developed,
and employed to calculate the structure of the outflow in the sub-slow magnetosonic 
region and the mass loading of the outflow, under conditions anticipated in 
the central engines of gamma-ray bursts.  The dependence of the mass flux
on the conditions in the disk, on magnetic field geometry, and on other factors 
is carefully examined for a range of neutrino luminosities expected in 
hyperaccreting black holes.  It is found that a fraction of up to a few percent of 
the neutrino luminosity produced inside the disk is deposited in the upper disk layers,
resulting in a steep rise of the specific entropy and the consequent
ejection of baryons along magnetic field lines.  For the range of conditions explored the 
final value of the dimensionless entropy per baryon in the wind is typically below 100. 
The fraction of neutrino
luminosity that is ultimately being converted to kinetic energy flux is shown to be 
a sensitive function of the effective neutrino 
temperature at the flow injection point, and the shape of magnetic field lines in the sub-slow 
region, but is practically independent of the strength of poloidal and toroidal 
magnetic fields.  We conclude that magnetic launching of ultra-relativistic polar outflows 
from the innermost parts of the disk is in principle possible provided the neutrino luminosity 
is sufficiently low, $L_\nu\simlt10^{52}$ erg s$^{-1}$ or so.  The conditions found to be optimal 
for the launching of an ultra-relativistic jet are also the conditions favorable for large 
neutron-to-proton ratio in the disk, suggesting that a large neutron excess in GRB jets, 
as often conjectured, may be possible.   However, the outflow time appears to be comparable 
to the neutronization timescale, implying that the electron fraction should evolve during 
the initial acceleration phase.  Further analysis is required to determine the composition 
profile in the wind.
\end{abstract}

\section{Introduction}

The central engines that power gamma-ray bursts (GRBs) are thought to
consist of a newly formed stellar mass black hole surrounded by a hot, dense, magnetized disk.  The 
prompt and afterglow emissions are inferred to be produced in ultra-relativistic outflows 
that are launched by the central engine and accelerate to Lorentz factors 
$\gamma_\infty>100$.  The free energy source of those relativistic, gamma-ray emitting jets can
be either the gravitational potential energy of accreted matter or the spin energy 
of a Kerr black hole.  Whether due to dissipation of some fraction of the spin
energy of the hole (van Putten \& Levinson 2003; Levinson 2005), or owing to hyper-accretion 
rates (Popham et al. 1999; Pruet et al. 2003), the disk surrounding the black 
hole is expected to be sustained at MeV temperatures, and its midplane density 
is expected to be high, in excess of $10^{10}$ gr cm$^{-3}$.  
Under such conditions the disk is optically thick to electromagnetic radiation 
and the dominant energy loss mechanism is neutrino emission.  Additionally,
the nuclear composition in the inner parts of the disk is likely to be neutron rich 
(Beloborodov, 2003; Pruet et al. 2003);  the neutron-to-proton ratio can in principle approach 30
if the disk material is cold enough ($T\simgt2$ MeV) and dense enough ($\rho\simgt 10^{11}$ 
gr cm$^{-3}$ at the disk midplane; see e.g., fig. 1 in Beloborodov [2003]).  
The prodigious neutrino luminosity is expected to drive a powerful wind from the disk.
This wind is most likely baryon rich and expands at sub or mildly relativistic speeds
except, perhaps, inside a core containing the putative baryon poor GRB jet.
The surrounding baryon rich wind may play an important role in the collimation of the central
jet (Levinson \& Eichler, 2000; Rosswog \& Ramirez-Ruiz, 2003), and its presence may also 
lead to some observational consequences (e.g., a supernova-like event).  
The nuclear composition of the disk outflow, particularly the central GRB jet, is 
an issue of considerable interest.  If, as often assumed, the GRB jet picks up nuclei from the 
disk (e.g., Derishev, Kocharovsky \& Kocharovsky 1999; Beloborodov 2003) then it may 
contain matter with a large neutron excess that is likely to affect its dynamics
(Fuller et al. 2000; Vlahakis, Peng \& Konigl 2003; Rossi et al. 2005), as well as some of the 
characteristics of the afterglow emission (e.g., Derishev, Kocharovsky \& Kocharovsky 
1999; Bachall \& Meszaros 2000).  In addition, the disk wind may be a potential site for 
nucleosynthesis (Pruet et al. 2004, 2005).
The nuclear composition of the disk outflow should depend on the conditions at the disk 
surface and the details of mass ejection process, particularly on the ratio of outflow time and 
neutronization timescale. 

The gross features of the physical picture outlined above are supported by recent numerical
simulations (e.g., Proga et al. 2003; McKinney, 2005a; and references therein), 
however, the detailed structure 
and properties of the polar outflow are only marginally resolved.  In particular, 
the origin of the fireball -  whether emanates from the inner regions of the disk or 
produced via a BZ mechanism, the nature of the baryon loading process, and the
nuclear composition of the outflow are yet open issues. Moreover, as will be shown 
below the mass ejection process is rather sensitive to the conditions at the flow 
injection point and the physical processes involved, which makes it difficult to 
study using global numerical simulations.

In order that a wind be accelerated hydrodynamically to high Lorentz factors, 
the entropy per baryon at the wind injection region must be extremely high.  Purely 
hydrodynamic fireballs demand
$s/k_B\simgt 10^5(\gamma_{\infty}/200)(kT_{d}/2MeV)^{-1}$ where $\gamma_{\infty}$ is the 
terminal Lorentz factor of the fireball and $T_d$ is the temperature at the base of the outflow.   
Although the specific entropy of the fireball should rise as it accelerates, owing to absorption 
of neutrinos escaping from the dense layers of the disk,
such extreme values are unlikely to be achieved.  The requirements on the baryon 
loading can be alleviated if the outflow is magnetically dominated.  In that case, 
the disk must support an ordered magnetic field with strength in excess of a few times $10^{14}$ G
to account for the observed GRB energies (e.g., Levinson \& Eichler 1993; Meszaros \& Rees, 1997).  
Such strong fields conceivably exist in the dense disks of hyperaccreting black holes 
(e.g.,  McKinney, 2005b).  Since the energy is extracted magnetically, acceleration of the 
outflow to high Lorentz factors does not require, in principle, 
high specific entropy, but rather a high magnetic energy per baryon $\cE_B$.
In the ideal MHD case $\cE_B>\gamma_{\infty}\sim 200$ and if the fireball accelerates
along magnetic field lines that are anchored to the disk, the question then arises as to how
baryon overloading of magnetic field lines can be avoided.   Vlahakis, Peng \& Konigl (2003) 
proposed that such 
high $\cE_B$ may not in fact be necessary if the ratio of baryon-to-magnetic flux is not 
conserved on magnetic flux surfaces.  Specifically, they considered a situation in which 
a neutron rich outflow, expelled along magnetic field lines that emanate from accreted disk 
material in the vicinity of the innermost stable circular orbit, is accelerated magnetically 
to a Lorentz factor of order 10, at which point the neutrons decouple.  The remaining
protons then continue to accelerate magnetically to $\gamma_\infty\sim200$.  
With n/p ratio of about 30, $\cE_B\sim10$ or so is sufficient before decoupling 
(but not smaller, otherwise the neutrons wont decouple at all).
However, such a high n/p ratio requires optimal conditions in the disk, and
essentially no evolution of the electron fraction $Y_e$ during the acceleration phase 
prior to decoupling, which is questionable.

The issue of mass loading has not been addressed in most of the previous work on 
magnetized, relativistic disk outflows.  The common approach, following 
the pioneering work of Blandford \& Payne (1982), is to seek self-similar
solutions of the trans-field equation (Li et al., 1992; Contopoulos 1994;
Vlahakis \& Konigle 2003).  While such a treatment considerably simplifies 
the analysis, it precludes the incorporation of gravity (in the relativistic case)
as well as neutrino heating in the flow equations, and in addition does not allow 
matching of the self-similar outflow solution
to a Keplerian disk (Li et al. 1992).  The immediate implication is that the 
self-similar outflow solutions cannot be extended down to the sub-slow magnetosonic region.
On the other hand, the mass-to-magnetic flux ratio is determined by the regularity condition
at the slow magnetosonic point, namely by the requirement that the outflow must pass 
smoothly through this point.  This means that solutions of the flow equations in the 
sub-slow magnetosonic region must be obtained in order for the mass flux to be calculated. 
Since the self-similar treatment is inapplicable in this region, one
must seek a different approach.  
In this paper we construct a model for GRMHD disk outflows with neutrino-driven mass ejection, and
employ it to calculate the structure of the flow in the sub-slow region.  Using the regularity 
condition at the slow magnetosonic point we carefully examine how the mass flux depends 
on the conditions at the injection point, on geometry of magnetic surfaces, and on other
wind quantities.  In \S 2 we present the basic equations for a GRMHD wind with external energy
and momentum sources.  In \S 3 we derive analytic expressions for the flow quantities in terms 
of the location and temperature at the slow point by employing the regularity condition,
and identify some systematic trends.  In \S 4 we present numerical
solutions.  Our approach is to integrate the flow equations assuming the magnetic field 
geometry is given. Since the slow point is located close to the disk midplane,
our choice of a particular field geometry reflects in part a choice of boundary conditions. 
We explore a range of field geometries in order to elucidate the dependence of the mass 
flux on the shape of the field lines.   

Neutrino-driven winds have been discussed previously in other contexts (e.g., Duncan et al. 1986;
Levinson \& Eichler 1993; Woosley et al. 1994; Witti et al. 1994; Qian \& Woosley 1996;
Thompson et al. 2001).  However, most previous models assume purely hydrodynamic, spherically 
symmetric winds, which are less relevant to the problem at hand.  A simplified model of
hydrodynamic disk outflow is presented in Pruet et al. (2004), mainly in connection with 
nucleosynthesis.  In this work we generalize 
those previous treatments to incorporate GRMHD effects and disk geometries.

\section{GRMHD Neutrino-Assisted Wind Model}

We consider a MHD wind expelled from the surface of a hot, magnetized disk surrounding a 
non-rotating black hole.  The range of conditions in the disk is envisioned to be
similar to that computed by Popham et. al. (1999) for hyperaccreting black holes
with mass accretion rates $10^{-1} - 10$  $M_{\sun}$ s$^{-1}$ and  viscosity parameters 
$\alpha_{vis}=0.01 - 0.1$.   Under such conditions the total neutrino luminosity emitted 
from the disk lies in the range $L_\nu=10^{51}$ - $10^{54}$ erg s$^{-1}$. 
We focus our attention on the inner regions of the disk
(within 10 Schwarzschild radii or so) where the midplane density exceeds about $10^9$ gr cm$^{-3}$, 
and where the major fraction of the neutrino luminosity is generated.

Now, for the range of temperatures considered the torus material should be a mixture 
of baryons and a light fluid (photons 
and electron-positron pairs in equilibrium).  The light and baryonic fluids will 
be tightly coupled in the sub-slow magnetosonic region, owing to the large Thomson depth.
Deep beneath its surface, 
the torus is in a hydrostatic equilibrium where the vertical gravitational force 
exerted on it by the black hole is supported by the baryon pressure.  In this region 
heating and cooling processes due to neutrino emission and absorption 
proceed at very high rates, and the temperature variation in the vertical direction
is expected to be very small.  As a consequence, the vertical density profile is roughly a Gaussian with a
rather small scale height ($h/R_0\simlt0.1 $; see below).  As the density drops, the ratio
of light fluid pressure to baryonic pressure increases roughly as $T^3/\rho$, until at
$\rho\simlt10^8 T_{MeV}^3$ gr cm$^{-3}$ the light fluid pressure dominates
over the baryonic pressure (see eq. [\ref{sigma}] below).  At this point hydrostatic equilibrium
can no longer be sustained and the matter starts accelerating along magnetic field lines.
This suggests that the mass flux carried by the wind should be
controlled by the heating and cooling processes at the base of the wind.  These processes
are therefore incorporated into the flow equations derived below.
To simplify the analysis we consider, in what follows, only time independent, axisymmetric
models.  As shown by Vlahakis \& Konigle (2003b), a relativistic MHD pulse can be adequately
described by the same equations.

\subsection{Basic equations}
Since, as explained above, the light and baryonic fluids will be tightly coupled  
at the base of the outflow, they can be treated as a single fluid.  We denote by  
$\rho$, $p$, $e$, $h=(e+p)/\rho c^2$, 
the proper baryon rest mass density, total pressure, total energy density, and dimensionless 
specific enthalpy of this mixed fluid.  The stress-energy tensor then takes the form:
\begin{equation}
T^{\alpha\beta} = h\rho c^2u^{\alpha}u^{\beta} + pg^{\alpha\beta}+{1\over4\pi}
(F^{\alpha\sigma}F^\beta_\sigma-{1\over4\pi}g^{\alpha\beta}F^2),
\label{T_M}
\end{equation}
where $u^{\alpha}$ is the four-velocity measured in units of c, $F_{\mu\nu}=\partial _\mu A_\nu
-\partial _\nu A_\mu$ is the electromagnetic tensor, and $g^{\alpha\beta}$ is the metric 
tensor.   In what follows we take spacetime to have a Schwarzschild geometry, defined by the 
line element
$ds^2=-\alpha^2dt^2+\alpha^{-2}dr^2+r^2d\theta^2+R^2d\phi^2$, with $\alpha^2
=1-r_s/r$, and $R=r\sin\theta$.
We denote by $q^{\beta}$ the source terms associated with energy and momentum transfer
by some external agent.  
The dynamics of the MHD system is then governed by the following set of equations: 
energy and momentum equations
\begin{equation}
\frac{1}{\sqrt{-g}}(\sqrt{-g}T^{\alpha\beta})_{,\alpha}+
\Gamma^{\beta}_{\ \mu\nu}T^{\mu\nu} = q^\beta, \label{Ttot=q}
\end{equation}
continuity equation
\begin{equation}
\frac{1}{\sqrt{-g}}(\sqrt{-g}\rho u^{\alpha})_{,\alpha}=0,
\label{continuity}
\end{equation}
and Maxwell's equations,
\begin{eqnarray}
F^{\beta\alpha}_{;\alpha}=\frac{1}{\sqrt{-g}}(\sqrt{-g}F^{\beta\alpha})_{,\alpha}
=4\pi j^{\beta},\label{F=j}\\
F_{\alpha\beta,\gamma}+F_{\beta\gamma,\alpha}+F_{\gamma\alpha,\beta}=0.\label{F=0}
\end{eqnarray}
We assume the flow to be stationary and axisymmetric ($\partial_t=\partial_\phi=0$),
and require infinite conductivity, $u^\alpha F_{\alpha\beta}=0$.
The above set of equations then admit two invariants: the mass-to-magnetic flux ratio 
\begin{equation}
\eta={\rho u_p\over B_p}
\label{eta}
\end{equation}
and the angular velocity,
\begin{equation}
\Omega=v^\phi-{\eta\over\sqrt{-g}\rho u^t}F_{r\theta}=v^\phi-{v_p\over B_p}{B_\phi\over R}.
\label{Omega}
\end{equation}
Here $u_p$ is the poloidal 4-velocity of the fluid defined by 
$u_p^2=u^ru_r+u^\theta u_\theta$, $v_p=u_p/\gamma$ is the corresponding 3-velocity, 
where $\gamma=\alpha u^t$ is the Lorentz factor, $v^\phi=u^\phi/u^t$,
$B_p$ is a rescaled poloidal magnetic field defined through
$B_p^2=(B^r B_r+B^\theta B_\theta)/\alpha^2$, and $B_\phi=rF^{r\theta}/\alpha$
is the toroidal magnetic field.
By contracting $u_\beta$ with eq. (\ref{Ttot=q}), using the identity 
$\Gamma^{\beta}_{\ \mu\nu}u_\beta u^\mu u^\nu=-u_\beta u^\nu(u^\beta)_{,\nu}$,
and the thermodynamic identity $dh=dp/\rho c^2+(k_B T/m_N c^2)ds$, where $s$ is the 
dimensionless specific entropy and $m_N$ is the nucleon rest mass, we obtain the change 
in $s$ along magnetic flux surfaces:
\begin{equation}
(\rho/m_N)k_BTu^\alpha s_{,\alpha}=-u_\alpha q^\alpha.
\label{s=q}
\end{equation}
By contracting $g_{\beta\gamma}$ with eq. (\ref{Ttot=q}) and taking the $t$ and $\phi$
components, we obtain
\begin{eqnarray}
\rho c^2 u^\alpha \cE_{,\alpha}=-q_t,\label{E=q}\\
\rho c^2 u^\alpha \cL_{,\alpha}=q_\phi,\label{L=q}
\end{eqnarray}
where
\begin{equation}
\label{E}
\cE=-h u_t -\frac{\sqrt{-g}}{4\pi\eta c^2}\Omega F^{r\theta}=
h\gamma\alpha-{\alpha R\Omega B_\phi\over 4\pi\eta c^2},\\
\end{equation}
and
\begin{equation}
\label{L}
\cL=h u_{\phi}-\frac{\sqrt{-g}}{4\pi\eta c^2} F^{r\theta}=
hu_\phi-{\alpha R B_\phi\over 4\pi\eta c^2},
\end{equation}
are the specific energy and angular momentum of the MHD system, respectively.
We henceforth assume that $q_\phi=0$, so that the specific angular momentum is conserved.
By employing eqs (\ref{eta})  (\ref{Omega})  (\ref{E})  (\ref{L}) we can express the toroidal 
magnetic field in terms of the flow parameters:
\begin{equation}
B_\phi=-{4\pi\eta\over \alpha R}\frac{\alpha^2 \cL-R
^2\Omega \cE}{(\alpha^2-R^2\Omega^2-M^2)}\label{Bphi},
\end{equation}
where $M$ is the Alfv\'en Mach number defined by $M^2=4\pi h\eta^2c^2/\rho=u_p^2/u_A^2$, 
with $u_A^2=B_p^2/(4\pi h \rho c^2)$.
Using eqs (\ref{Omega}), (\ref{E}) and (\ref{L}), and the normalization condition
$u^\alpha u_\alpha=-1$ yields
\begin{equation}
1+u_p^2={(\cE-\Omega \cL)^2(\alpha^2-R^2\Omega^2-2M^2)-
[(\cL/R)^2-(\cE/\alpha)^2]M^4
\over h^2(\alpha^2-R^2\Omega^2-M^2)^2}.
\label{Berno}
\end{equation}
We can differentiate eq. (\ref{Berno}) along a given stream line to obtain
\begin{equation}
(\ln u_p)^\prime={N+N_q\over D},
\label{motion}
\end{equation}
where ($^\prime$) denotes derivative along the stream line $\Psi=$const, and
\begin{eqnarray}
D=-(\alpha^2-R^2\Omega^2-M^2)^2(u_p^2-u_{SM}^2)(u_p^2-u_{FM}^2)/u_A^2,\label{D}\\
N=\zeta_1(\ln B_p)^\prime+\zeta_2(\ln \alpha)^\prime+\zeta_3(\ln R)^\prime,\label{N}\\
N_q=\zeta_4(\ln \cE)^\prime+\zeta_5 (\ln s)^\prime.\label{Nq}
\end{eqnarray}
Here $u_{SM}$ and $u_{FM}$ are the slow and fast magnetosonic wave speeds, respectively, and 
are given in eqs. (\ref{uSM}) and (\ref{uFM}), and $c_s$ is the sound 4-velocity defined 
by $c_s^2=a_s^2/(1-a_s^2)$ with $a_s^2$ given in
eq. (\ref{as2}) below.  The coefficients $\zeta_i$ are functions of the flow parameters, 
viz., $\zeta_i=\zeta_i(\Omega,\cL,\cE,s,M)$, and are derived in the appendix. 
Equation (\ref{motion}) generalizes the result derived by Takahashi et al. (1990) to 
the non-adiabatic case.  As seen, energy and momentum exchange with an external
agent (i.e., $N_q\ne0$) formally modifies the conditions at the critical points.  However,
in practice we find this to be a small correction.  As shown below, the main effect of heating 
is to enhance the temperature and the specific entropy at the slow point.
In cases where the stream function is known, the above set of equations, augmented by an
equation of state $h=h(\rho,s)$, can be solved to yield the structure of the flow.
The poloidal magnetic field can then be expressed in terms of the stream function.  
Taking $A_\phi$ as the stream function $\Psi(r,\theta)$ we obtain,
\begin{equation}
B_p^2=\frac{r^2\alpha^2(\Psi_{,r})^2+(\Psi_{,\theta})^2}{R^2r^2\alpha^2}.
\label{Bp-2}
\end{equation}

\subsection{Equation of state}
For the range of densities and temperatures considered here, we find that the relativistic 
particles in the vicinity of the slow magnetosonic point are non-degenerate. 
The total kinetic pressure of the mixed fluid is then given by $p=p_l+p_b$, where 
\begin{equation}
p_l ={11\over 12}aT^4= 1.2\times10^{26}T_{MeV}^4, \ \ \ {\rm dyn}\  {\rm cm^{-2}}
\label{pl}
\end{equation}
is the light fluid pressure, and
\begin{equation}
p_b=\rho {k T\over m_N}\simeq 9\times10^{26}\rho_9T_{MeV} \ \ \ {\rm dyn}\  {\rm cm^{-2}}
\label{pb}
\end{equation}
is the pressure contributed by the baryons, with $T_{MeV}$ being the temperature 
in MeV units, and $\rho_9$ the rest mass
density in units of $10^9$ g  cm$^{-3}$.
The dimensionless enthalpy per baryon of the mixed fluid is given by 
\begin{equation}
h=1+{4p_l\over \rho c^2}+{5\over2}{p_b\over\rho c^2}.
\label{enthalpy}
\end{equation}
It is convenient to define the thermodynamic quantity
\begin{equation}
\sigma={4p_l\over p_b}=0.53{T_{MeV}^3\over\rho_9}.
\label{sigma}
\end{equation}
Clearly, the pressure is dominated by the light fluid when $\sigma >4$. 
In terms of $\sigma$ the change in entropy is related to the change in temperature
and density through
\begin{equation}
ds={m_Nc^2\over kT}\left(dh-{dp\over \rho c^2}\right)=\left({3\over2}+3\sigma\right)
{dT\over T}-(1+\sigma){d\rho\over \rho}.
\label{entropy}
\end{equation}
It is readily seen from eqs (\ref{sigma}) and (\ref{entropy}) that $\sigma$ is 
roughly the entropy per baryon in regions where the light fluid pressure dominates; 
that is $s\simeq\sigma$ in the limit $\sigma>>1$.  
The relativistic sound speed can now be expressed as:
\begin{equation}
a_s^2=\left(\frac{\partial \ln h}{\partial \ln \rho}\right)_s={p_b\over \rho c^2h}
\frac{5+10\sigma+2\sigma^2}{3(1+2\sigma)}.
\label{as2}
\end{equation}
We also need the change in $h$ at constant density.  Using the above results we obtain
\begin{equation}
\left(\frac{\partial h}{\partial s}\right)_\rho={kT\over m_Nc^2}\frac{5+8\sigma}{3(1+2\sigma)}.
\label{an2}
\end{equation}

\subsection{Rates for neutrino heating and cooling}
The neutrino flux emitted from the dense disk layers depends on the 
structure of and conditions in the disk, which are uncertain.  Quite 
generally the neutrino luminosity comes predominantly from disk radii 
within a few $r_s$ (Popham et al. 1999).  To
simplify our calculations, we invoke a spherical neutrino source
with a radius $R_\nu=10^6R_{\nu6}$ cm and a total luminosity
$L_{\nu}=10^{53}L_{\nu53}$ erg s$^{-1}$, which are treated as free parameters.
Some fraction of the energy of the escaping neutrinos is deposited in the surface 
layers of the disk, via pair neutrino annihilation into electron - positron                         
pairs and neutrino capture on neutrons and protons, thereby giving rise to                     
a significant heating of the matter near the injection point of the wind. 
It is conceivable that turbulence or magnetic field dissipation may provide
additional heating of the surface layers, but these processes are difficult to
model and will be ignored.
Detailed account of the various neutrino heating and cooling process is given 
in Qian \& Woosley (1996).  The heating rate due to neutrino absorption is
approximately (see also Bethe and Wilson 1985)                      
\begin{equation}
\dot Q_{\nu n} \simeq  5\times                                                             
10^{31}\frac{L^{3/2}_{\nu53}}{R^3_{\nu6}}\rho_9f_{\nu n}(r) \ \
{\rm erg\ cm^{-3}\ s^{-1}}, 
\label{Qnun}
\end{equation} 
and that due to neutrino-antineutrino annihilation into electron-positron pairs
is (e.g., Goodman et al. 1987; Qian \& Woosley 1996)
\begin{equation}
\dot Q_{\nu\bar{\nu}}                                                                                
\simeq10^{31}\frac{L_{\nu 53}^2}{R_{\nu6}^4} f_{\nu\nu}(r)\ \ 
{\rm erg\ cm^{-3}\ s^{-1}}.  
\end{equation}
Here
$f_{\nu n}(r)=[1-(1-R_\nu^2/r^2)^{1/2}$], and 
$f_{\nu\nu}(r)=f_{\nu n}^4(r)[1-R^2_\nu/r^2+4(1-R_\nu^2/r^2)^{1/2}+5]$.  
Energy  gain by neutrino capture on nucleons dominates at densities above 
\begin{equation}                       
\rho_9 = 0.2 (L_{\nu53}/R_{\nu6}^2)^{1/2}(f_{\nu\nu}/f_{\nu n}). 
\label{rho-1}                                  
\end{equation}   
As a result of neutrino heating, the surface layers of the disk                         
will quickly rise to temperatures in excess of several Mev,
at which emission of secondary neutrinos by the inverse processes;
electron and positron capture on nucleons       
\begin{equation}
\epsilon_{\nu n}\simeq 10^{27}\rho_9 T_{MeV}^6\ {\rm erg\ s^{-1}\ cm^{-3}},
\label{epsnun}
\end{equation}                         
and electron-positron annihilation into neutrinos           
\begin{equation}
\epsilon _{\nu\bar{\nu}} = 5\times 10^{24} T_{MeV}^9\ {\rm ergs\ s^{-1}\                         
cm^{-3}},
\end{equation}
becomes the dominant energy loss mechanism.  Equating the last two rates one                     
obtains the density below which the pair neutrino cooling rate exceeds the                      
cooling rate due to electron and positron capture on nucleons, 
\begin{equation}                            
\rho_9 = 5\times 10^{-3}T^3_{MeV}.   
\label{rho-2}
\end{equation}
The critical density is found to be always well above this value, implying that 
URCA cooling is the dominant energy loss mechanism in the regions of 
interest.  The source term associated with energy transfer is given by
\begin{equation}                            
q^t= \dot Q_{\nu n} +\dot Q_{\nu\bar{\nu}}-\epsilon_{\nu n}-\epsilon_{\nu \nu}
\label{q^t}
\end{equation}
Well below the slow-magnetosonic point, where the flow velocity is small and adiabatic cooling
is negligible, the temperature profile is determined the condition $q^0=0$.  In this region the 
density exceeds the values given by eqs (\ref{rho-1}) and (\ref{rho-2}), 
and so the heating and cooling rates are dominated by capture (eqs. [\ref{Qnun}] and [\ref{epsnun}]).
The temperature profile is then given to a good approximation by
\begin{equation}                            
T_{MeV}(r)=6\left({L_{\nu53}\over R^2_{\nu6}}\right)^{1/4}f_{\nu n}^{1/6}. 
\label{Tprof}  
\end{equation}

\section{The Sub-Slow Magnetosonic Region - General Considerations}

\subsection{The flow near the disk surface}
Deep beneath the slow magnetosonic point, the mater is close to hydrostatic 
equilibrium.  Heating and cooling by the neutrinos proceeds at high rates and
the temperature is maintained at the level given by eq. (\ref{Tprof}).  If the pressure there 
is dominated by the baryons, then the density scale height at radius $R_0$ 
is roughly $h/R_0\sim (kTR_0/GMm_p)^{1/2}\simeq10^{-1.5}(R_0/r_s)^{1/2}T_{MeV}^{1/2}$, 
which is typically much smaller than unity near the black hole. Consequently, the base of
the flow is located close to the disk midplane.
Moreover, the ratio of the sound and Alfv\'en speeds is $c_s/u_A\simeq 10^{-1.5}T_{MeV}^2/B_{p15}$, so
that under the conditions envisioned
the Alfv\'en Mach number is anticipated to be very small in the sub-slow region.  To 
order $O(M^2)$ eqs. (\ref{TakA})- (\ref{TakC}) yield:
\begin{eqnarray}
\zeta_1=-(\alpha^2-R^2\Omega^2)^3(1+u_p^2)c_s^2,\label{zet1}\\
\zeta_2=-\frac{\alpha^2(\alpha^2-R^2\Omega^2)(\cE-\Omega\cL)^2}{(1-a_s^2)h^2},\\
\zeta_3=-\zeta_2{R^2\Omega^2\over\alpha^2}.\label{zet3}
\end{eqnarray}
Consider now a field line described by the equation $R=
R(z)$, and denote $R_0=R(z=0)$, $R_0^\prime=dR/dz|_{z=0}$, and $\tilde{N}=(dz/dl)^{-1} N$, where
$d/dl=(dz/dl)d/dz$ is the derivative along the field line, and $^\prime$ denotes now differentiation 
with respect to $z$.  Close to the disk 
midplane we can expand $R(z)$ about 
$z=0$.  To first order in $z$ eq. (\ref{N}) yields
\begin{equation}
\tilde{N}=\zeta_1(\ln B_p)^\prime+{\zeta_2\over\alpha^2}\left\{(\Omega^2_k-\Omega^2)
[R_0R_0^\prime+(R_0^\prime)^2 z+
R_0R_0^{\prime\prime}z]+
\Omega^2_k[1-3(R_0^\prime)^2]z\right\}, 
\label{N-base}
\end{equation} 
where $\Omega_k$ denotes the Keplerian angular velocity at $R=
R_0$, viz., $\Omega_k^2=r_s/2R_0^3$.
At the base of the flow adiabatic losses are negligibly small and $N_q\simeq0$.
Now, in reality the poloidal velocity of the outflow should vanish at some height $z_d(R_0)$
above the disk where the streamlines of inflowing matter joins the outflow.  
At this effective surface the angular velocity $\Omega$ is expected to be equal to 
the local Keplerian angular velocity.
To leading order we then find $\Omega^2=\Omega_k^2 (1-3R_0^\prime z_d/R_0)$, where the 
radius at which the field line meets the surface is related to $R_0$ through 
$R_d\simeq R_0+R_0^\prime z_d $.  Substituting the latter result into eq. (\ref{N-base}) we obtain
\begin{equation}
\tilde{N}=\zeta_1(\ln B_p)^\prime+{\zeta_2\over\alpha^2}
\Omega^2_k[z-3(R_0^\prime)^2(z-z_d)]. 
\label{N-d}
\end{equation} 
From eq. (\ref{D}) it is evident that $D<0$ when $u_p<u_{SM}$.  Consequently,
the condition $N<0$ must be fulfilled in the sub-slow region in order that the flow 
be accelerated.  If the magnetic field lines are diverging, viz., $(\ln B_p)^\prime<0$, as one 
might expect, then the first term on the right hand side of eq. (\ref{N-d}) is always
positive, implying that the second term must be negative in the sub-slow region.  
As seen, there are two distinct cases.  If the magnetic field lines are inclined at 
an angle larger than $\pi/6$ to the vertical ($R_0^\prime>1/\sqrt{3}$), then the second 
term on the right hand side of eq. (\ref{N-d}) 
has a root at $z_1=[3(R_0^\prime)^2/(3(R_0^\prime)^2-1)]z_d$ and it changes sign across.
Along such field lines the outflow is centrifugally driven and can be initiated even 
in the cold fluid limit (at which $\zeta_1=0$), as was first shown by 
Blandford \& Payne (1982).  The slow magnetosonic point is 
located close to the disk surface, at $z_{sm}\le z_1$ (the exact location depends on the value of the 
sound speed at the disk surface).  
If the magnetic field lines are inclined at an angle smaller than $\pi/6$ to the vertical,
then the second term in eq. (\ref{N-d}) is negative at $z=0$ and decreases linearly with $z$ in the 
region where the above expansion holds.  This corresponds to the regime of stable equilibrium, as
defined in Blandford \& Payne (1982).  Along such field lines the mass flux is thermally 
driven, similar to the case of a spherical wind.  As shown below, the slow magnetosonic 
point in this case is located on larger scales, $z_{sm}\sim R_0$, where the sound speed roughly equals
the escape velocity.  The above derivation is in accord with the results obtained by Ogilvie (1997) 
\footnote{Note though that 
the term $\zeta_1(\ln B_p)^{\prime}$ is neglected in eq. (5.35) of Ogilvie, which may be justified
near the disk surface.  However, the omission of this term precludes treatment of transonic 
outflows along low inclination field lines}. 

\subsection{Conditions at the slow-magnetosonic point}
At the critical point $D=0$ and hence $N+N_q=0$.  Neglecting terms of order $M^2$
and higher, one finds $u_{SM}=c_s$ and  
\begin{equation}
(1+c_s^2)a_s^2(\ln B_p)_c^\prime+{\gamma_c^2\over \alpha^2_c}(\alpha_c\alpha_c^\prime
-R_cR_c^\prime\Omega^2)-\gamma_c{q^t\over 3\rho c^3u_p}=0,
\label{crit-cond}
\end{equation}
where eqs (\ref{zet1})-(\ref{zet3}) and (\ref{N_q})-(\ref{zet5}) have been 
used.  Henceforth, subscript $c$ refers to quantities at the critical point.
Equation (\ref{crit-cond}) can be solved now for the sound speed $a_s$.
The result can be simplified further by noting that at the slow magnetosonic point
the flow is sub-relativistic, so that $c_s^2\simeq a^2_s<<1$, and $\gamma_c\simeq1$.
Moreover, at the slow point the last term on the LHS of eq. (\ref{crit-cond}) is 
approximately $q^t/ 3\rho c^3u_p\simeq 10^{-9.5}(L_{\nu53}/R_{\nu6}^2)^{3/2}
(R_\nu/r_c)^2a_s^{-1}$ cm$^{-1}$, where eq. (\ref{Qnun}) has been employed, and
can be neglected to lowest order.  
With the above approximations the solution of eq. (\ref{crit-cond}) reads:
\begin{equation}
a_s^2=a_{sc}^2= -{(\alpha_c\alpha_c^\prime-R_cR_c^\prime\Omega^2)
\over \alpha_c^2(\ln B_p)_c^\prime}.
\label{as2-crit}
\end{equation}
Now, in general we expect $(\ln B_p)_c^\prime=-A/R_c$, where $A$ is an order unity number,
with $A\simeq2$ for conical flux tubes, $A=0$ for cylindrical flux tubes, and $A\simeq1$
in the case of a self-similar magnetic field.  In terms of $\Omega_k$ 
eq. (\ref{as2-crit}) can be expressed as:
\begin{equation}
a_{sc}^2={R_c\over A}\left\{\frac{R_c R^\prime_c(\Omega_k^2 R_0^3-\Omega^2r_c^3)
+\Omega_k^2z_cR_0^3}{r_c^2(r_c-r_s)}\right\},
\label{as2-crit2}
\end{equation}
where $r^2_c=R^2_c+z^2_c$. From the last equation it is evident that $a_s^2<<1$.  Below
we find that typically $a_s$ lies in the range $0.1 -0.01$.

\subsubsection{Centrifugally driven mass flow}

Since in this case the critical point occurs near the disk surface we can expand $R_c=R(z_c)$ 
about $z=0$ as before.  Equation (\ref{as2-crit2}) then reduces to
\begin{equation}
a_{sc}^2\simeq {r_s\over R_0+r_s}{z_c+3(R_0^\prime)^2(z_d-z_c)\over A R_0}.
\label{as-cent}
\end{equation}
Adopting for illustration $A=2$, $r_s/R_0=0.3$, $z_d/R_0=0.1$, we estimate $a_{sc}^2<10^{-2}$
at the critical point.
Since the latter occurs close to the surface, adiabatic cooling is unimportant
and the temperature profile is given to a good approximation by eq. (\ref{Tprof}).
For the range of neutrino luminosities considered here we expect $T_{c}\simgt 1$ MeV. 
The mass loading of the outflow, $\rho_c a_{sc}$, would depend on the density at the 
flow injection point.  However, the critical density and, hence, the mass flux 
cannot be arbitrarily small.  To estimate the minimum mass flux, we
employ eq. (\ref{as2}) to write
\begin{equation}
a_{s}^2=10^{-3}T_{MeV}\frac{5+10\sigma+2\sigma^2}{3(1+2\sigma)},
\label{as-cent2}
\end{equation}
and conclude that under the conditions envisioned $\sigma$ cannot be much larger than unity at 
the critical point in order for eq. (\ref{as-cent}) to be satisfied.  Equation (\ref{sigma})
then implies a critical density of $\rho_c\simgt 10^8 T_{cMeV}^3$ g cm$^{-3}$, and a mass flux 
$\dot{M}\simeq \rho_c a_s \pi R_0^2 > 10^{30}$ g s$^{-1}$, for $R_0=r_s=10^{6.5}$ cm.  
Consequently, outflows along field lines having inclination angles larger than $\pi/6$ 
are expected to be sub-relativistic, owing to the large mass flux driven from the surface.
Numerical integration of the full equations confirms this result.

\subsubsection{Thermally driven mass flow}
In this case the critical point occurs higher above the disk.  As will be shown in sec. 4, 
the specific entropy rises steeply during the initial acceleration phase, so that 
$\sigma>>1$ near the critical point.
Using eqs. (\ref{as2}) and (\ref{as2-crit}) we can relate the critical density 
to the critical temperature. In the limit $\sigma>>1$ we obtain
\begin{equation}
\rho_c\simeq 10^5 {T_{cMeV}^4\over a_{sc}^2}
\ \ \ {\rm gr \ cm^{-3}}.
\label{rho-crit}
\end{equation}
As seen, the critical density is a sensitive function of the critical temperature and, therefore,
adiabatic cooling can largely suppress the mass flux.
An estimate of the mass-to-magnetic flux ratio can be obtained by noting  
that the critical point is located roughly at $z_c\sim R_0$,
so that $B_{pc}\simeq B_{p0}$.  One then finds
\begin{equation}
\eta=\frac{\rho_c c_s}{B_{pc}}\simeq 3 {T_{cMeV}^4\over B_{p015}}a_{sc}^{-1}
 \ \ {\rm gr \ cm^{-2} \ s^{-1} \ G^{-1}}.
\label{eta-crit}
\end{equation}
From the above result we further estimate that for the range of conditions considered here 
the relativistic energy per baryon of the matter is dominated by the nucleon 
rest mass, implying $h_c\gamma_c\alpha_c\simeq1$.  To a good approximation the total 
energy per baryon (eq. [\ref{E}]) is then $\cE=1+\cE_B$, with the magnetic energy per baryon given by
\begin{equation}
\cE_B\simeq 5\times 10^2 T_{cMeV}^{-4}B_{p015}^2 a_{sc}
\left({-B_\phi\over B_p}\right)_0.
\label{E-ant}
\end{equation}
At the highest neutrino luminosities adiabatic cooling can be neglected even 
along low inclination field lines, and the 
temperature at the slow point is given to a good approximation by eq. (\ref{Tprof})
with $r=r_c$.  The above quantities can then be estimated once the location of the 
slow point is known.  At lower neutrino luminosities, however, adiabatic cooling will 
suppress the critical temperature, rendering such estimates highly 
uncertain.  One must then integrate the flow equations in order to accurately compute 
the location of and temperature at the slow point.  This is discussed in the next section.

Equations (\ref{as2-crit}) and (\ref{rho-crit}) 
imply that, to order $O(M^2)$, the critical mass flux on a given field 
line, $\rho_c c_s$, depends only on the gravitational potential,
the angular velocity, the neutrino luminosity (through the critical temperature $T_{cMeV}$),
and geometry of streamlines, and is independent of the strength of poloidal and toroidal
magnetic fields at the disk surface.  Consequently, for a given choice of $L_{\nu}$ and magnetic 
field geometry, there is a critical value of $B_{p0}B_{\phi 0}$ above which 
$\cE_B>1$ and the flow is essentially magnetically driven.  Numerical integration 
of the flow equation confirms this conclusion, as shown below.

\section{Numerical Model} 
The model outlined above is characterized by three parameters: the black hole mass $r_s$,
the neutrino luminosity $L_{\nu}$, and the neutrinospheric radius $R_{\nu}$.
Fixing these parameters, eqs. (\ref{s=q}), (\ref{E=q}) and (\ref{motion}) can 
be solved simultaneously once the magnetic field geometry is known.
In general, however, the stream function is unknown a priori, and one must 
solve the highly non-linear trans-field equation simultaneously with eqs 
(\ref{s=q}), (\ref{E=q}), (\ref{enthalpy}) and (\ref{entropy}), and subject to
appropriate boundary conditions to fully determine the flow structure. 
Unfortunately, the inclusion of gravity and external heating breaks scale-freeness,
and precludes separation of variables as in self-similar treatments.  This
means that a self-consistent calculation of the field geometry is impractical.  
Since in this work we are merely interested in estimating the mass flux driven by 
neutrino heating it is sufficient to integrate the above flow equations only up to the 
slow magnetosonic point.  As will be shown below, the latter is located rather close 
to the disk surface where the field geometry should anyhow reflect the boundary
conditions.  Our approach shall be to specify the stream function and 
use it to integrate the above set of flow equations along a given 
field line.  We explore different
field geometries in order to elucidate the dependence of baryon loading on the 
basic characteristics of the flux surfaces.
The system under consideration has three integrals of motion: The angular velocity $\Omega$,
the angular momentum $\cL$, and the mass-to-magnetic flux ratio, $\eta$.  
In our approach, $\Omega$, $\cL$, and the values of $\cE$ and $B_{p}$ at the origin are 
treated as additional free parameters, and $\eta$ as an eigenvalue of the system.

In what follows we first describe the magnetic field geometries used in the our calculations. 

\subsection{Topology of magnetic surfaces}

\subsubsection{split monopole}
In our first example we consider a split monopole field, with its center shifted a distance $a$ 
from the black hole along the hole rotation axis (see Fig 1).  In cylindrical coordinates,
the field line equation reads: $R(z)=\kappa(z+a)=\kappa z+R_0$, 
where $\kappa=\tan\tilde{\theta}$ is the field 
line parameter and $R_0=\kappa a$.
The stream function has the form $\Psi=\Psi_0(1-\cos\tilde{\theta})$.  Substituting the latter into eq. 
(\ref{Bp-2}) yields the poloidal magnetic field:  
\begin{equation}
\left({B_p\over B_{p0}}\right)^2={\alpha^2_0R_0^4\over \alpha^2 R^4}{\left\{(1+\kappa^2)
-r_sR_0^2 (R^2+z^2)^{-3/2}\right\}\over (1+\kappa^2)-r_s/R_0},
\label{Bp-sm}
\end{equation}
where $B_{p0}=B_p(z=0)$.  Note that this geometry implies a current sheet at the disk 
midplane.

The proper length of a given stream line, measured in units of $R_0$, is given by
\begin{equation}
dl={1\over R_0}(g_{rr}dr^2+g_{\theta\theta}d\theta^2)^{1/2}={1\over\alpha}\left[1+\kappa^2
-{r_s R_0^2\over (R^2+z^2)^{3/2}}\right]^{1/2}{dz\over R_0},
\end{equation}
and the derivative along stream lines by,
\begin{equation}
u^\alpha\partial_\alpha={u_p\over R_0}{d\over dl}=u_p \alpha\left[1+\kappa^2
-{r_s R_0^2\over (R^2+z^2)^{3/2}}\right]^{-1/2}{d\over dz}.
\label{prop-split}
\end{equation}
From eqs. (\ref{E=q}) and (\ref{prop-split}) we find that the change in specific energy per 
unit length along a given streamline is 
\begin{equation}
{d\cE\over dl}={q^tR_0\alpha\over \rho c^3 u_p}.
\label{E=q-split}
\end{equation}
Under the assumption that the neutrino trajectories are radial we obtain
\begin{equation}
u_\nu q^\nu=-q^t\alpha\left\{\gamma-u_p\left(\alpha {R_0dl\over dz}\right)^{-1}
{\kappa R+z\over \sqrt{R^2+z^2}}\right\}.
\end{equation}
Substituting the last equation into eq. (\ref{s=q}) yields
\begin{equation}
{ds\over dl}={q^tR_0\alpha\over kT(\rho/m_N) u_p}
\left\{\gamma-u_p\left(\alpha {dl\over dz}\right)^{-1}
{\kappa R+z\over \sqrt{R^2+z^2}}\right\},
\end{equation}
which in the non-relativistic limit, $\gamma=\alpha=1$, reduces to the 
result derived by Qian \& Woosley (1996).

\subsubsection{Self-similar geometry}

In case of an outflow from a disk, we assume that the streamlines in the vicinity of 
some radius $R=R_0$ can be described by 
\begin{equation}
(R,z)=R_0[g(\xi),\xi],
\label{streamlines}
\end{equation}
subject to the boundary condition $g(\xi=0)=1$.  The stream function is the solution
of the equation $r\sin\theta/R_0(\Psi)=g(r\cos\theta/R_0(\Psi))$.  
Using eq. (\ref{Bp-2}) one finds
\begin{equation}
\left({B_p\over B_{p0}}\right)^2={\alpha_0^2 \over[\alpha_0^2+(g_{\xi 0})^2] g^2\alpha^2}
\left\{{1+(g_\xi)^2\over (g-\xi g_\xi)^2}-{r_s\over R_0(g^2+\xi^2)^{3/2}}\right\},
\label{Bp-disk}
\end{equation}
where subscript $0$ refers to the value of the corresponding quantity at the the 
disk midplane, viz., at $\xi=0$,
and $g_\xi=d g/d\xi$.  
To obtain the derivative along streamlines we note that in terms of the proper 
length of a given streamline, $R_0dl=(g_{rr}dr^2+g_{\theta\theta}d\theta^2)^{1/2}$, we have
\begin{equation}
u^\alpha\partial_\alpha=u_p d/dl=u_p(dl/d\xi)^{-1}d/d\xi,
\end{equation}
with
\begin{equation}
(dl/d\xi)^2={1\over\alpha^2}
\left\{1+(g_\xi)^2-{r_s\over R_0}
{(g-\xi g_\xi)^2\over(g^2+\xi^2)^{3/2}}\right\}.
\label{arc}
\end{equation}
For the spherical neutrino source invoked above  we obtain,
\begin{equation}
u_\nu q^\nu=-q^t\alpha\left\{\gamma-u_p\left(\alpha {dl\over d\xi}\right)^{-1}
{gg_\xi+\xi\over \sqrt{g^2+\xi^2}}\right\},
\end{equation}
and
\begin{eqnarray}
{d\cE\over d\xi}={q^t R_0\alpha\over \rho c^3 u_p}{dl\over d\xi},\label{E=q-disk}\\
{ds\over d\xi}={q^tR_0\over kT(\rho/m_N) u_p}
\left\{\gamma\left(\alpha {dl\over d\xi}\right)-u_p{gg_\xi+\xi\over 
\sqrt{g^2+\xi^2}}\right\}. \label{s=q-disk}
\end{eqnarray}

\subsection{Boundary Conditions and Numerical Integration}

Equations (\ref{s=q}), (\ref{E=q}), (\ref{Bphi}), (\ref{motion})-(\ref{Nq}),
(\ref{enthalpy})-(\ref{as2}), and (\ref{q^t})
have been integrated numerically for different field geometries. 
In each run we fix the parameters $B_{p0}$, $B_{\phi0}$, $\Omega$, $r_s/R_0$, 
$R_\nu/R_0$ and $L_{\nu53}/R_{\nu6}^2$.   In 
what follows, we find it more convenient to 
parametrize the neutrino flux in terms of the effective temperature 
$T_\nu\simeq 6 (L_{\nu53}/R^2_{\nu6})^{1/4}$ MeV.
The temperature at the origin ($z=0$) is determined by the condition 
$q^t(z=0)=0$, and is given to a very good approximation by 
$T_0\simeq T_\nu f_{\nu n}^{1/6}$ (see  eqs. [\ref{Qnun}] and [\ref{Tprof}]) .
The integration starts at sufficiently dense disk layers where the pressure 
is marginally dominated by the baryon pressure.  To be more concrete, for a given $T_0$ 
the density at the origin $\rho_0$ was chosen such that $\sigma\simlt 4$. 
The entropy per baryon at the origin is taken to be 
$s=8.7+\ln(T_{MeV}^{3/2}/\rho_{9})+\sigma$ (Popham et al. 1999).
The parameter $\eta$ is then adjusted iteratively by changing the boundary value of the poloidal 
velocity $u_{p0}$, until a smooth transition across the slow magnetosonic
point is obtained.  To verify that the eigenvalue $\eta$ is insensitive to $\rho_0$ \footnote{The eigenvalue $\eta$
should depend, to some extent, on $\rho_0$ through the condition $q^t(z=0)=0$ that fixes the 
the temperature at the disk layer where $\rho=\rho_0$.}, 
each integration has been repeated several times for a given choice 
of our model parameters, each time with a different value of $\rho_0$.  
We find, indeed, that the dependence of $\eta$ on $\rho_0$
is very weak provided that the corresponding value of $\sigma$ is not too large.

\section{Results}
We now present solutions for the flow structure in the sub-slow region, for the 
two classes of field geometries described above.  For the models calculated using
the self-similar field geometry we invoked a parabolic field-line shape:
\begin{equation}
g(\xi)=1+\delta \xi^2,
\label{parabolic}
\end{equation}
where $0<\delta<1$ is an additional parameter, independent of $R_0$, that  
controls the divergence of magnetic field lines above the disk.  Note that
for this choice of $g$ $(\ln B_p)^\prime=0$ at $z=0$.  All the results presented
below were computed using a black hole mass $M_{BH}=3M_{\sun}$, and neutrinospheric 
radius $R_{\nu}=3r_s=10^{6.5}$ cm.  For this choice of 
$R_{\nu}$ the effective temperature $T_\nu$ 
lies in the range $1$ to $6$ MeV for the range of neutrino luminosities discussed in 
Popham et al (1999).  Sample results and the parameters employed are summarized in 
figs 2 - 8 and in table 1.

Typical solutions are shown in fig 2, where the change along a field line of 
the flow quantities indicated is plotted against the normalized height above the disk 
midplane, $z/R_0$, for the split monopole geometry with $\kappa=0.2$, $R_0=R_\nu$,
$B_{p0}=10^{15}$ G, $(-B_{\phi}/B_p)_0=0.1$, and $\Omega=\Omega_k$, and for 
different values of the neutrino luminosity given in terms of $T_\nu$. 
For each of the cases shown the temperature at the origin is $T_0=T_\nu$ since
$R_0=R_\nu$. 
As seen, the slow magnetosonic point is located rather close to the disk surface,
at $z\simeq R_0$, as one might expect.
Well beneath the slow magnetosonic point the flow velocity is small, the density is high,
and the specific entropy is dominated by the baryons, that is $\sigma\simlt 4$.  In this region
the net energy deposition rate nearly vanishes, viz., $q^t\simeq0$, and the temperature profile 
is given to a good approximation by equation (\ref{Tprof}).  As the flow accelerates its 
temperature starts falling, and since the neutrino 
cooling rate is very sensitive to the temperature $q^t$ increases rapidly leading to a 
steep rise of the entropy per baryon to its terminal value, as seen in fig 2.  
The reason why the energy deposition per unit length 
peaks well below the slow point, as seen, is that
$u_p$ there is small and, hence, the time over which a fluid element is exposed to the
external neutrino flux, $dt\sim dl/u_p$, is large.
Although formally the total energy per baryon is not conserved along streamlines,
we find that it changes only slightly (to less than 0.1 \%) as the flow accelerates.
This is because the net energy deposited per baryon comprises only a small fraction
of the baryon rest mass.   The main effect of neutrino heating is not to change the 
net outflow energy along streamlines, but rather to enhance the baryon load by 
increasing the light fluid pressure in layers of higher baryon density.  
Thus, in practice the total energy per baryon can be considered conserved in the 
relativistic and  mildly relativistic cases (this may not be true if the wind is highly
sub-relativistic), and so the value of $\cE$ presents, essentially, an upper limit
for the asymptotic Lorentz factor of the wind.  The values of $\cE$, $\eta$ and the mass flux,
which we somewhat arbitrarily define as $\dot{M}=\rho_0u_{p0}2\pi R_0^2$ (for a two-sided
outflow), that corresponds to the cases depicted in fig. 2 are listed in table 1. 
The values of $\eta$ and $\cE$ computed numerically are in good agreement with 
analytic expressions given in eqs. (\ref{eta-crit}) and (\ref{E-ant}).

In fig. 3 we plotted solutions obtained for a centrifugally driven outflow along
field lines in the unstable regime ($\kappa>1/\sqrt{3}$).  The effective disk surface
in this example is located at $z_d=0.1 R_0$.  The temperature at 
the surface is $T_0=2$ MeV, and the remaining parameters are the same as in fig. 2.
As seen, the slow magnetosonic point is located very near the surface.
The correspoding mass fluxes are $\dot{M}/(10^{30}\ {\rm gr\ s^{-1}})=7.6$, $76$, $95$,
and $101$ for $\kappa=1$, 2, 3, and 4, respectively.   In addition to this example, 
we also explored other regimes of the parameter space. The main conclusion 
is that, quite generally, the mass loading of centrifugally driven winds from neutrino cooled disks 
is too large to allow acceleration to relativistic speeds. Since we are merely interested in 
relativistic winds here
we focus, in what follows, on the regime $\kappa <1/\sqrt{3}$.

Fig 4 exhibits the dependence of the total energy per baryon $\cE$ on the ratio 
$(-B_\phi/B_p)_0$.  It is clearly seen that the magnetic energy per baryon, $\cE_B=\cE-1$,
depends linearly on $(-B_\phi/B_p)_0$, in agreement with the 
analytic expression given in eq. (\ref{E-ant}).  The ratio of mass-to-magnetic
flux $\eta$ is found to be practically independent of $(-B_\phi/B_p)_0$.
In the examples depicted
in fig 4, $\eta$ changes by less than 3\% over the range  $(-B_\phi/B_p)_0=10^{-3} - 1$.
In fact we find that the mass flux $\rho u_p$ is highly insensitive also to the strength 
of the poloidal field $B_{p0}$,
provided the Alfv\'en Mach number $M$ is sufficiently small at the slow magnetosonic point
(that is, $c_s<<u_A$ there), and that it depends predominantly on the neutrino flux, the
geometry of magnetic field lines, and the angular velocity $\Omega$.  This confirms that for 
a given choice of the latter parameters, $\cE_B$ is proportional to the 
product $(-B_pB_\phi)_0$.  The results presented in fig 4 can then be readily 
rescaled for other choices of $B_{p0}$.  

The dependence of baryon loading on the inclination angle of the field line is examined
in fig 5, where $\eta$ is plotted against the parameter $\kappa$ for a 
fixed $R_0$ (which corresponds to changing the distance $a$ in fig 1 keeping $R_0$ fixed).
The corresponding mass flux, $\dot{M}=\rho_0u_{p0}2\pi R_0^2$, is indicated on the right
axis.  As expected, the mass flux is 
larger on field lines with larger inclination angles, owing to the slingshot effect.
The dependence of mass loading on the radius at which the field line intersects the disk
is demonstrated in fig. 6, where 
$\eta$ and $\dot{M}$ are plotted against $R_0$ for a fixed $\kappa$.  As 
seen, the trend is that the mass flux increases with increasing 
$R_0$.  This mainly reflects the fact that the escape velocity is smaller on field 
lines that emanate from larger disk radii $R_0$.  However, the dependence is weaker 
than one might anticipate, owing to the decrease in the heating rate with 
increasing $R_0/R_\nu$ (see eq. [\ref{Qnun}] and below).   We also made some runs 
with different values of $r_s/R_\nu$ and found the mass flux to be sensitive 
also to this parameter at low $T_\nu$.  General relativistic effects appear to be important
near the black hole, but we do not attempt here to quantify them.

For completeness, we also examined the dependence of $\eta$ on the angular velocity of
magnetic flux tubes.  Although the latter is naively expected to be nearly Keplerian, it
is conceivable that various effects may give rise to non-Keplerian rotation of some 
flux tubes during certain periods.  The general trend, as seen in fig. 7, is for the baryon 
load to increase with increasing $\Omega$, by virtue of the larger contribution 
of centrifugal forces.  However, a change in $\Omega$ may also lead to a change
in the Poynting flux, so the implications for the asymptotic Lorentz factor are not
straightforward.

Fig 8 presents results obtained using the self-similar configuration 
with $g(\xi)$ given by eq. (\ref{parabolic}), for $\delta=0.5$ (left panels)
and $\delta=0.2$ (right panels).   The behavior of the solutions is rather similar to the 
split monopole case.  We generally find low mass fluxes for which $\cE>>1$ 
on field lines that emanate from within several $r_s$ and that diverge not too fast
at low enough neutrino temperatures ($T_\nu\simlt2$ MeV).
We have also checked other forms of $g(\xi)$.  In all cases we find that at $T_\nu\simlt2$
MeV relativistic outflows can be launched from the innermost disk regions.

We finally address the question whether the nuclear composition of the expelled matter
changes during the acceleration of the flow.   Significant evolution of the electron fraction
$Y_e$ is expected if the neutronization timescale, $t_n\simeq10 T_{MeV}^{-5}$ s, is much shorter 
than the outflow time $t_{d}=(R_0/c)\int{dl/u_p}$.   The evolution of $Y_e$ should 
take place at the base of the flow, where the temperature is highest and, hence, $t_n$ is shortest,
and where $u_p$ is smallest and, hence, $t_d$ is longest.  For most cases studied above we find
the outflow time to be comparable to $t_n$, and so the n/p ratio is expected to evolve during
the initial acceleration phase.   Quantitative treatment of the nuclear processes in the wind 
is beyond the scope of this paper and is left for a future investigation.

\section{Conclusions}
We have explored stationary, axisymmetric GRMHD disk outflows in Schwarzschild geometry 
with neutrino-driven mass ejection, under conditions anticipated in hyperacrreting
black holes.  We have examined the dependence of mass loading of the outflow
on the neutrino luminosity emitted by the disk and on other wind quantities,
for a range of effective neutrino temperatures $T_\nu =
1-6$ MeV, which for the simplified geometry of the neutrino emission zone invoked 
in our model corresponds to a neutrino luminosity range $L_\nu =10^{51} - 10^{53}$
erg s$^{-1}$.  We find that up to a few percent of 
the neutrino luminosity produced inside the disk is deposited at the upper disk layers 
and used up to eject baryons along magnetic field lines.
The neutrino-driven mass flux depends predominantly on the effective neutrino 
temperature at the base of the flow, on magnetic field geometry in 
the sub-slow magnetosonic region, and on the angular velocity of magnetic flux
surfaces, but is highly insensitive to the strength of poloidal and toroidal 
magnetic field components near the surface provided the Alfv\'en Mach number is sufficiently small 
at the slow point; the dependence on neutrino luminosity and the shape 
of magnetic field lines near the disk surface appears to be particularly sensitive. 
The heating of the wind by the escaping neutrinos results in a steep rise 
of the entropy per baryon, $s$, during the initial acceleration phase, after which it 
saturates.  The final value of $s$ is larger for lower $L_\nu$, but does not seem to reach
extreme values.  For the range of conditions explored above it is typically below 
$100 k_B$ per baryon. 

Our principle conclusion is that ejection of relativistic outflows from the 
innermost disk radii, within several $r_s$ or so, is possible in principle for certain magnetic field 
configurations even in non-rotating black holes, provided the neutrino luminosity is 
sufficiently low, and the magnetic field is sufficiently strong . 
In the case of the split monopole field delineated in section 4.1,
we obtain a mass flux of $\rho u_p\sim 10^{14}$ gr cm$^{-2}$ 
s$^{-1}$ on field lines in the vicinity of the innermost stable circular orbit
that make an angle $\tilde{\theta}\sim12^\circ$ with the 
rotation axis, for a neutrino luminosity $L_\nu\sim10^{52}$  erg s$^{-1}$ (
corresponding to $T_\nu\sim2$ MeV).  The outflow along those field
lines can, therefore, accelerate to Lorentz factors $\gamma_\infty\simlt100$ if the magnetic field 
strength at the disk surface is $B_{p0}\sim 10^{15}$ G.  The mass flux depends sensitively
on both $\tilde{\theta}$ and $T_\nu$, as can be seen from figure 5 and table 1, and so a steep profile
of the Lorentz factor across the polar jet is naively expected.  We find a similar trend also 
for the other field geometries adopted.  Thus, the picture often envisioned, 
of an ultra-relativistic core surrounded by a slower, baryon rich wind 
(e.g., Levinson \& Eichler 1993, 2000;  ) seems a natural consequence of neutrino-assisted 
MHD disk outflows.   At higher effective temperatures, $T_\nu>2$ MeV or so, the baryon load 
is typically well in excess of  the value inferred in GRB fireballs, and the wind is 
expected to be slow.  This conclusion may be altered if the central black hole is 
rapidly rotating. 

The hyperaccretion process is likely to be intermittent, leading to temporal changes
in the neutrino luminosity.  This should result in large variations of the Lorentz
factors of consecutive fluid shells expelled from the disk in the polar region, owing to 
the sensitive dependence of mass loading on $L_\nu$.  In this situation we expect strong shocks
to form in the outflow.  If the polar disk outflow is associated with the GRB-producing 
jet, then the observed gamma-ray emission can be quite efficiently produced behind those shocks.
   
It is worth noting that the conditions we find to be optimal for launching an ultrarelativistic jet
in the polar region, are also the conditions favorable for large neutron-to-proton ratio in the disk.
For example, by employing the results presented in 
table 2 of Popham et al. (1999), we estimate $T_\nu<2$ MeV for accretion rates 
$\dot{M}_{acc}\simlt 0.1 M_{\sun}$ s$^{-1}$ into
a 3 $M_{\sun}$  Schwarzschild black hole.   Under such conditions a neutron-to-proton 
ratio larger than 10 is expected inside the disk within a few $r_s$ for viscosity parameters 
$\alpha<0.03$ (Pruet et al. 2003; see also Beloborodoc 2003).  If turbulent mixing can 
quickly lift up neutron rich matter to the surface layers, then in can be picked up 
by the outflow.  Beloborodov (2003) has shown that deneutronization is not expected at
temperatures below about $8$ MeV if the mixing timescale is of the order of $\Omega_k^{-1}$. 
However, because at low $T_\nu$ the flow time is comparable to the 
neutronization timescale, the electron fraction $Y_e$ may evolve 
as the flow accelerates and the final nuclear composition of the outflow may somewhat change.
Further analysis is needed to determine the wind composition, but
the trend seems to be that ultrarelativistic disk outflows
can plausibly contain neutron rich matter.   This is in accord with the 
conjecture often made, that fireballs pick up neutron rich material from the 
disk.

I thank Arieh Konigle for inspiring conversations, the referee Nektarios Vlahakis for useful comments,
and Jason Pruet for helpful correspondence.
This work was supported by an ISF grant for the Israeli Center for High Energy Astrophysics

\appendix
\section{Equation of Motion for General Relativistic, Neutrino-Assisted MHD Wind}
The projection of the momentum equation on the poloidal direction yields eq. (\ref{motion}).
We can equivalently derive this equation by differentiating eq. (\ref{Berno}) along a 
given streamline $\Psi=$ const (e.g., Camenzind 1986).   To simplify the notation let us denote
\begin{eqnarray}
k_0=\alpha^2-R^2\Omega^2,\label{k0}\\
k_2=(\cE-\Omega \cL)^2,\label{k2}\\
k_4={\cL^2\over R^2}-{\cE^2\over\alpha^2}.\label{k4}
\end{eqnarray}
Differentiating the above equations along a stream line, making use of the fact that $\eta$, $\Omega$
and $\cL$ are conserved on magnetic surfaces, yields
\begin{eqnarray}
(k_0)^\prime=2\alpha^2(\ln\alpha)^\prime-2R^2\Omega^2(\ln R)^\prime,\\
(k_2)^\prime=2(\cE-\Omega \cL)\cE^\prime,\label{k2prime}\\
(k_4)^\prime={2\cE^2\over\alpha^2}(\ln \alpha)^\prime -2{\cL^2\over 
R^2}(\ln R)^\prime 
-2{\cE^2\over \alpha^2}(\ln \cE)^\prime.\label{Ek4prim}
\end{eqnarray}
Taking the enthalpy to be a function of the density $\rho$ and the entropy per baryon $s$ we further obtain,
\begin{eqnarray}
(\ln h)^\prime=a_s^2(\ln \rho)^\prime +{\partial \ln h\over \partial \ln s}(\ln s)^\prime,
\label{lnh}\\
(\ln M^2)^\prime=(\ln h)^\prime-(\ln \rho)^\prime=(a_s^2-1)(\ln \rho)^\prime
+{\partial \ln h\over \partial \ln s}(\ln s)^\prime,
\label{lnM2}
\end{eqnarray}
where $a_s^2=(\partial\ln h/\partial \ln \rho)_s$ is the adiabatic sound speed.
Upon differentiating eq. (\ref{Berno}) along a stream line, using the above results together
with eqs. (\ref{as2}) and (\ref{an2}), we arrive at
\begin{equation}
(\ln u_p)^\prime=(N+N_q)/D,
\label{aligned-q}
\end{equation}
where the terms in the nominator are given by
\begin{equation}
N=\zeta_1(\ln B_p)^\prime+\zeta_2(\ln \alpha)^\prime+\zeta_3(\ln R)^\prime,
\label{Napp}
\end{equation}
with
\begin{eqnarray}
\zeta_1=-(k_0-M^2)^2\left[(1+u_p^2)(k_0-M^2)c_s^2-M^2{B_\phi^2\over4\pi h\rho}\right],\label{TakA}\\
\zeta_2=\frac{1}{h^2(1-a_s^2)}\left\{{M^6\cE^2\over\alpha^2}-
\left(3\cE^2-{R^2\Omega^2\cE^2\over\alpha^2}
-{2\cL^2\over R^2}\right)M^4+\alpha^2k_2(3M^2-k_0)\right\},\label{TakB}\\
\zeta_3=\frac{1}{h^2(1-a_s^2)}\left\{-{\cL^2\over R^2}M^6
+\left(3\cL^2\Omega^2-{\alpha^2\cL^2\over R^2}-{2R^2\Omega^2
\cE^2\over \alpha^2}\right)M^4 -R^2\Omega^2k_2(3M^2-k_0)\right\},\label{TakC}
\end{eqnarray}
and 
\begin{equation}
N_q=\zeta_4(\ln\cE)^\prime+\zeta_5 (\ln s)^\prime,
\label{N_q}
\end{equation}
with
\begin{eqnarray}
\zeta_4={1\over h^2(1-a_s^2)}(k_0-M^2)[(k_0-2M^2)
(\cE-\Omega\cL)\cE+M^4\cE^2/\alpha^2],\label{zet4}\\
\zeta_5={(5+8\sigma)\over (5+10\sigma+2\sigma^2)}{s c_s^2\over h^2}
[k_4M^6-k_2(k_0^2-3k_0M^2+3M^4)], \label{zet5}
\end{eqnarray}
and the denominator by
\begin{equation}
D=(\alpha^2-R^2\Omega^2-M^2)^2\left[(u_p^2-c_s^2)(\alpha^2-R^2\Omega^2-M^2)
+M^2{B_\phi^2\over4\pi h\rho}\right].
\end{equation}
In terms of the slow and fast magnetosonic speeds,
\begin{eqnarray}
u_{SM}^2=K-\sqrt{K^2-c_s^2u_A^2(\alpha^2-R^2\Omega^2)},\label{uSM}\\
u_{FM}^2=K+\sqrt{K^2-c_s^2u_A^2(\alpha^2-R^2\Omega^2)},\label{uFM}
\end{eqnarray}
where 
\begin{equation}
K={1\over2}\left[(\alpha^2-R^2\Omega^2)u_A^2+c_s^2+{B_\phi^2\over4\pi h\rho}\right],
\end{equation}
the denominator $D$ can be rewritten as in eq. (\ref{D}).

\clearpage
\begin{table}
\caption{\baselineskip=11pt  Numerical models}
\begin{center}
\begin{tabular}{|ccccccccccc|} \hline\hline
Model\tablenotemark{a} & ($\kappa$, $\delta$)\tablenotemark{b}& $R_0/r_s$ 
&$R_0/R_\nu$& 
$L_{\nu53}/R_{\nu6}^2$ &$T_\nu$   & $T_0$ &$\eta$ & $\dot{M}$ &$s_\infty$\tablenotemark{c}&$\cE$  \\ 
   &      &     &       &      & MeV & MeV &gr cm$^{-2}$ s$^{-1}$G$^{-1}$ & $10^{30}$gr s$^{-1}$ &   &     \\ \hline
SM1...& 0.2 & 3      &   1       &  0.012    &2        & 2   &0.18    & $0.011$ & 44  & $492$  \\  
SM2...& 0.2  &3      &   1       &  0.062    &3        & 3  &7.4    & $0.47         $ & 32  &  $13$ \\
SM3...& 0.2  &3      &   1       &  0.197    &4        & 4 &116     & $7.4 $ & 29  &  $1.8$  \\ 
SM4...& 0.1  &3      &   1       &  0.012    &2        & 2 &0.028    & $0.0018 $ & 87  &  $3100$  \\ 
SM5...& 0.1  &3      &   1       &  0.197    &4        & 4 &22   & $1.39           $ & 55  &  $5.1$  \\ 
SM6...& 0.3  &3      &   1       &  0.012    &2           & 2 &1.6   & $0.1           $ & 19  &  $50$  \\ 
SM7...& 0.3  &3      &   1       &  0.197    &4           & 4 &1368   & $86.6        $ & 15  &  $1.03$  \\ 
SM8...& 0.2  &9      &  3     &  0.012       &2        & 1.25 &0.37   & $0.024 $ & 18  &  $161$  \\ 
SM9...& 0.2  &9      &  3     &  0.197       &4        & 2.5  &396    & $25$              & 15   &  $1.1$  \\ 
SS1...& 0.5  &3      &   1       &  0.012   &2            & 2  &26.4     & $1.67$             & 14    &  $4.3$  \\ 
SS2...& 0.5  &3      &   1       &  0.062   &3            & 3  &1008    & $64$              & 13     &  $1.05$  \\ 
SS3...& 0.2  &3      &   1       &  0.012   &2            & 2  &0.29    & $0.018$         & 29       &  $309$  \\
SS4...& 0.2  &3      &   1       &  0.062   &3            & 3  &313     & $19.8$           & 13     &  $1.2$  \\ \hline
\end{tabular}
\tablenotetext{a}{
``SM'' stands for split monopole geometry, ``SS'' for self-similar geometry.}
\tablenotetext{b}{
The numerical values in this column correspond to values of $\kappa$ in the ``SM'' models,
and to values of $\delta$ in the ``SS'' models (see text for further details).}
\tablenotetext{c}{$s_\infty$ refers to the asymptotic value of the dimensionless entropy per baryon $s$.}
\end{center}
\end{table}
 
\clearpage
\begin{figure}[f1]
\plotone{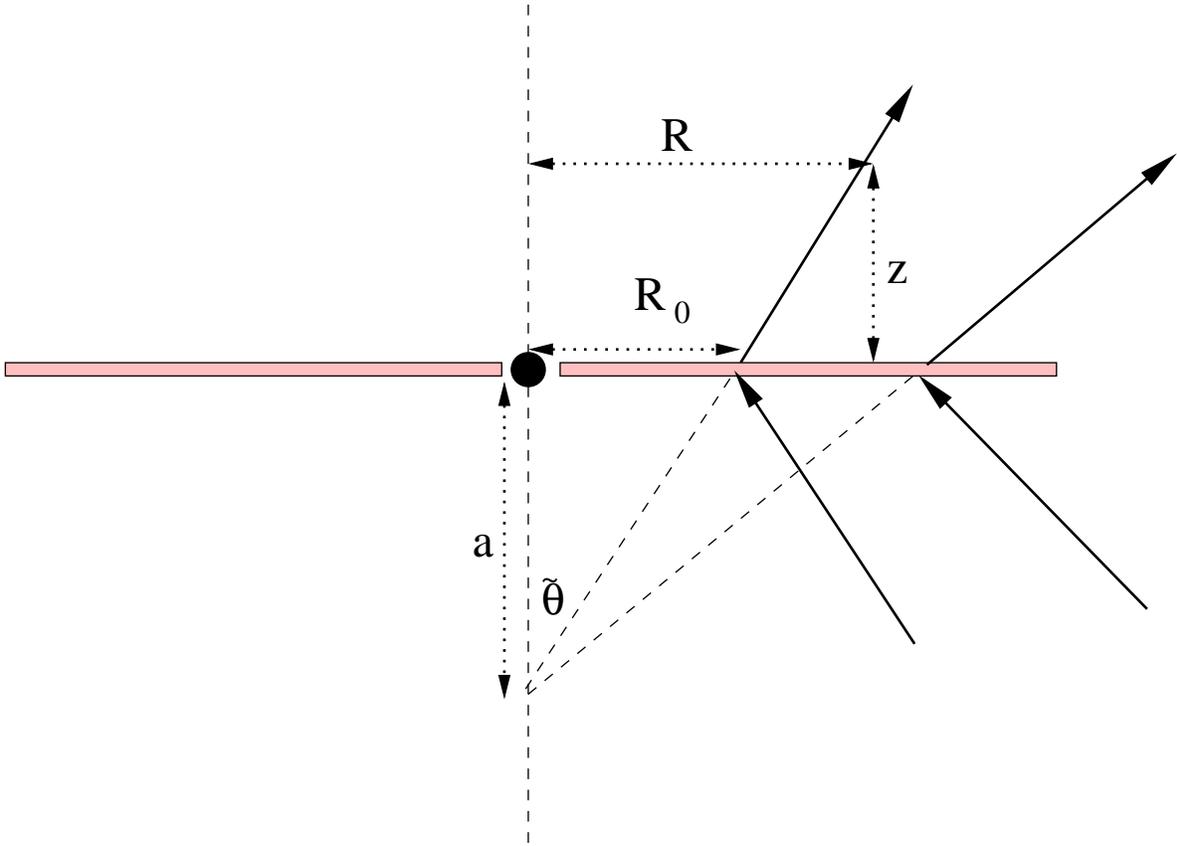}
\caption{Sketch of the split monopole configuration employed for the numerical computations.
The origin of the monopole is shifted a distance $a$ from the black hole 
along the symmetry axis of the system.  The magnetic field is supported 
by a current sheet in the equatorial plane.  The field line parameter
is defined as $\kappa=\tan\tilde{\theta}$. The radius $R_0$ at which the field line meets
the disk is indicated.}
\label{f1}
\end{figure}

\clearpage
\begin{figure}[f2]
\plotone{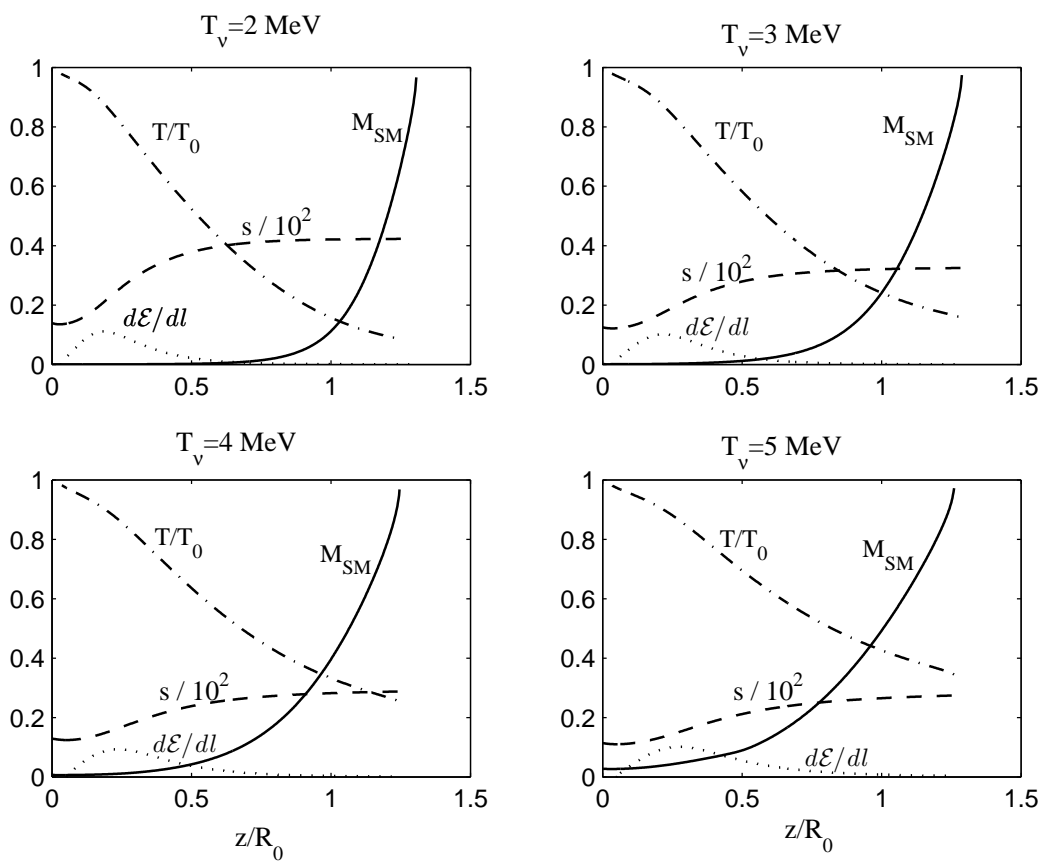}
\caption{Profiles of various quantities in the 
sub-slow magnetosonic region, computed using the split monopole magnetic field of fig. 1
with $\kappa=0.2$, $B_{p0}=10^{15}$ G, $(-B_\phi/B_{p})_0=0.1$, and Keplerian rotation, viz.,
$\Omega=\Omega_k$.  Each panel corresponds to a run with a different neutrino 
luminosity (indicated in terms of the effective 
temperature $T_\nu$; see text for details).  
The quantities plotted in each panel are: the slow magnetosonic Mach number $M_{SM}$ (solid line),
the dimensionless entropy per baryon $s$ (dashed line), the temperature $T$ in units of the initial 
temperature $T_0$ (dotted-dashed line), and the energy deposition per baryon per unit 
length along the streamline measured in units of $m_pc^2/R_0$, $d\cE/dl$ (dotted line).
All quantities are given as functions of the normalized height above the disk midplane $z/R_0$.  
}
\label{f2}
\end{figure}

\clearpage
\begin{figure}[f3]
\plotone{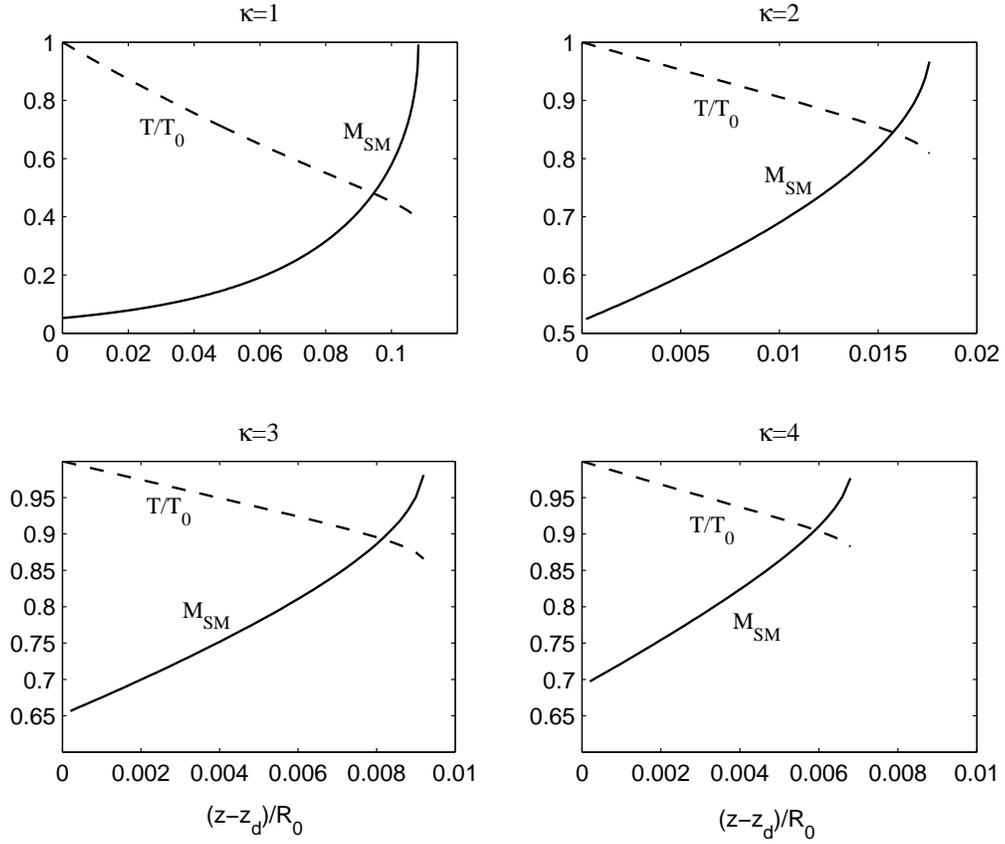}
\caption{Profiles of $T/T_0$ and $M_{SM}$ for centrifugally driven outflows.
The surface temperature in all cases shown is $T_0=2$ MeV, and the remaining parameters
are the same as in fig. 2. }
\label{f3}
\end{figure}

\clearpage
\begin{figure}[f4]
\plotone{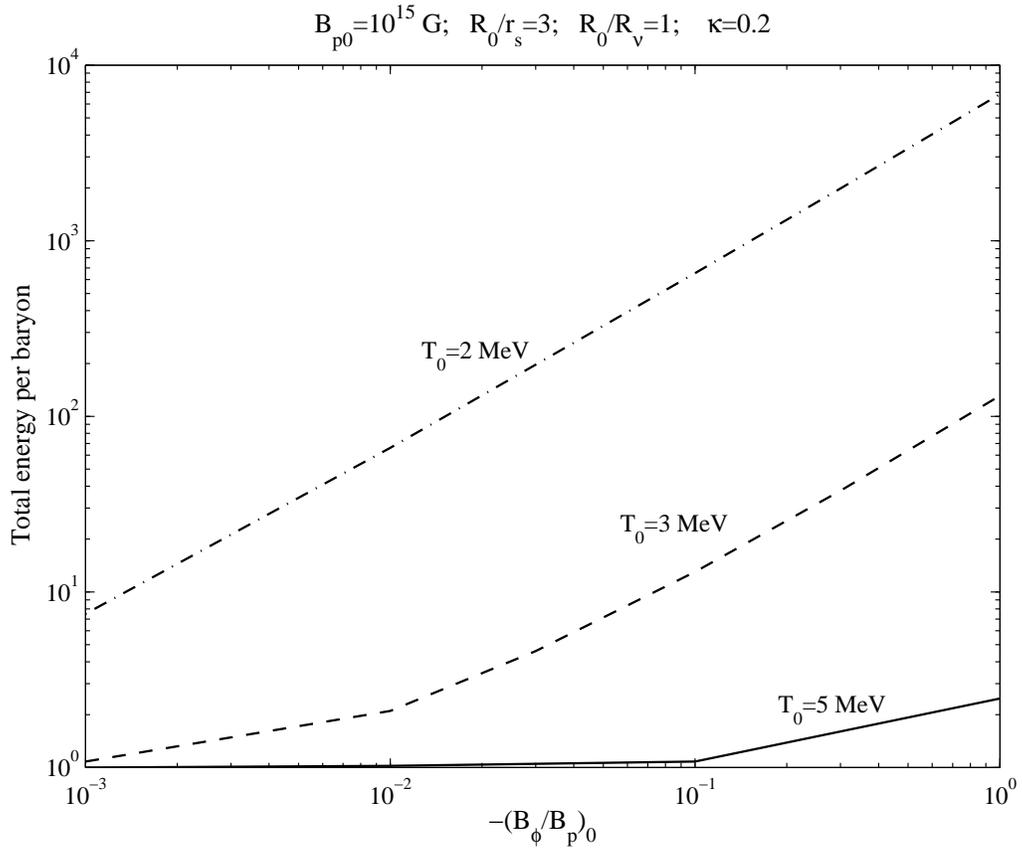}
\caption{Total energy per baryon, $\cE$, versus the ratio of the toroidal and poloidal components of the 
magnetic field at the disk surface, $-(B_{\phi}/B_p)_0$.}
\label{f4}
\end{figure}

\clearpage
\begin{figure}[f5]
\plotone{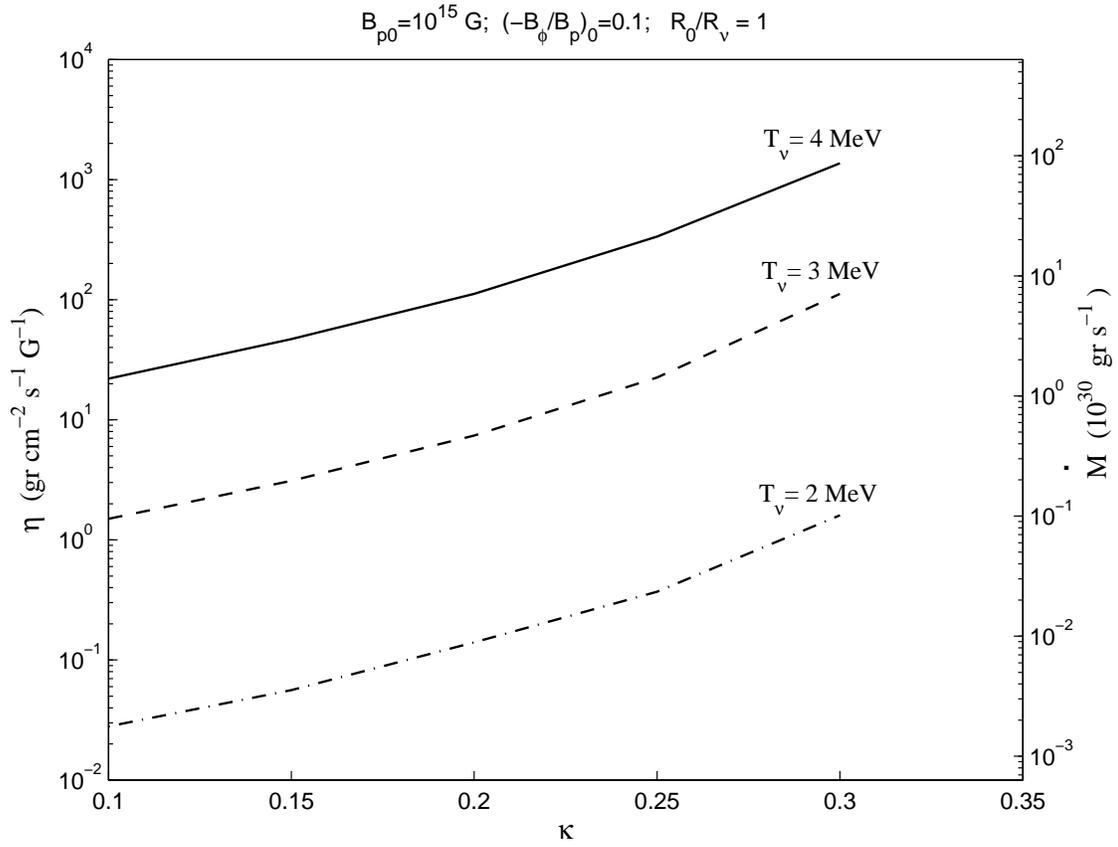}
\caption{Mass-to-magnetic flux ratio $\eta$ (left axis) as a function of the split-monopole 
field line parameter $\kappa$.  The corresponding values of the 
mass flux $\dot{M}$, as defined in the text, are indicated on the right axis.}
\label{f5}
\end{figure}

\clearpage
\begin{figure}[f6]
\plotone{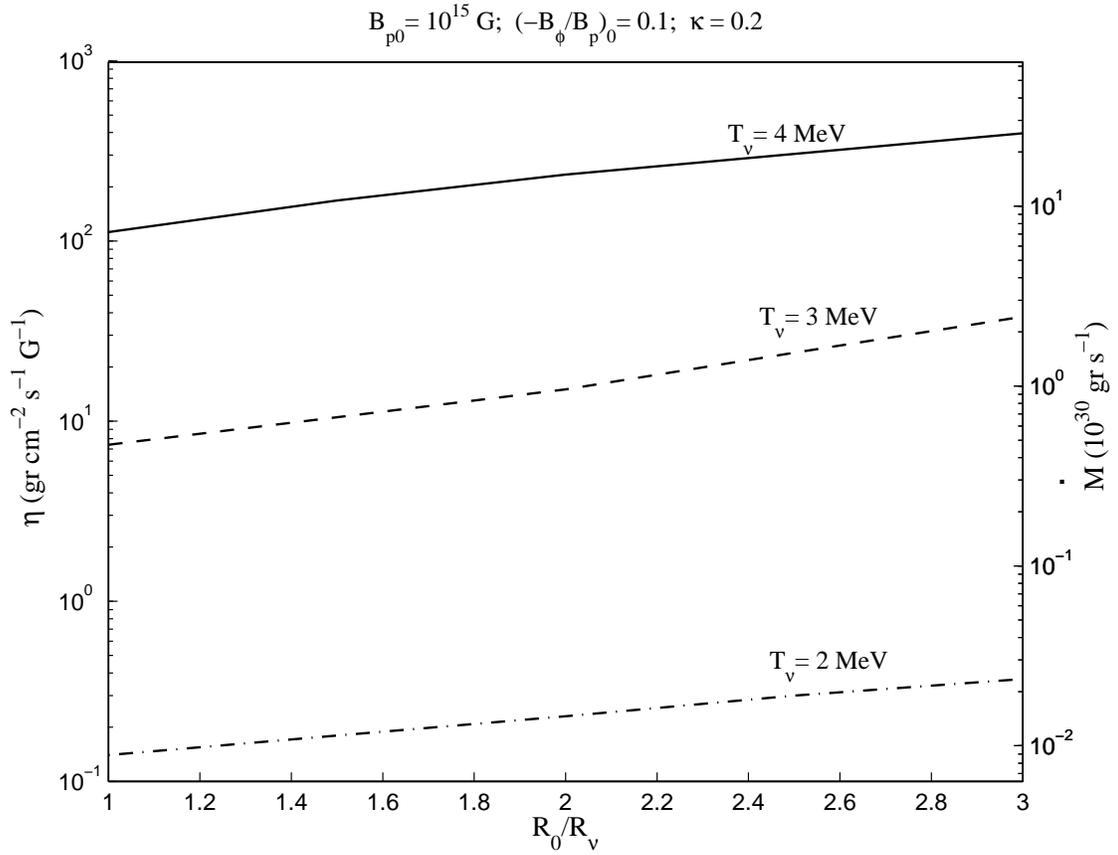}
\caption{The dependence of $\eta$ and $\dot{M}$ on the radius $R_0$
at which the field line intersects the disk (see fig. 1). The distance $a$ in fig. 1
is adjusted such that $\kappa$ is kept fixed as $R_0$ is varied}
\label{f6}
\end{figure}

\clearpage
\begin{figure}[f7]
\plotone{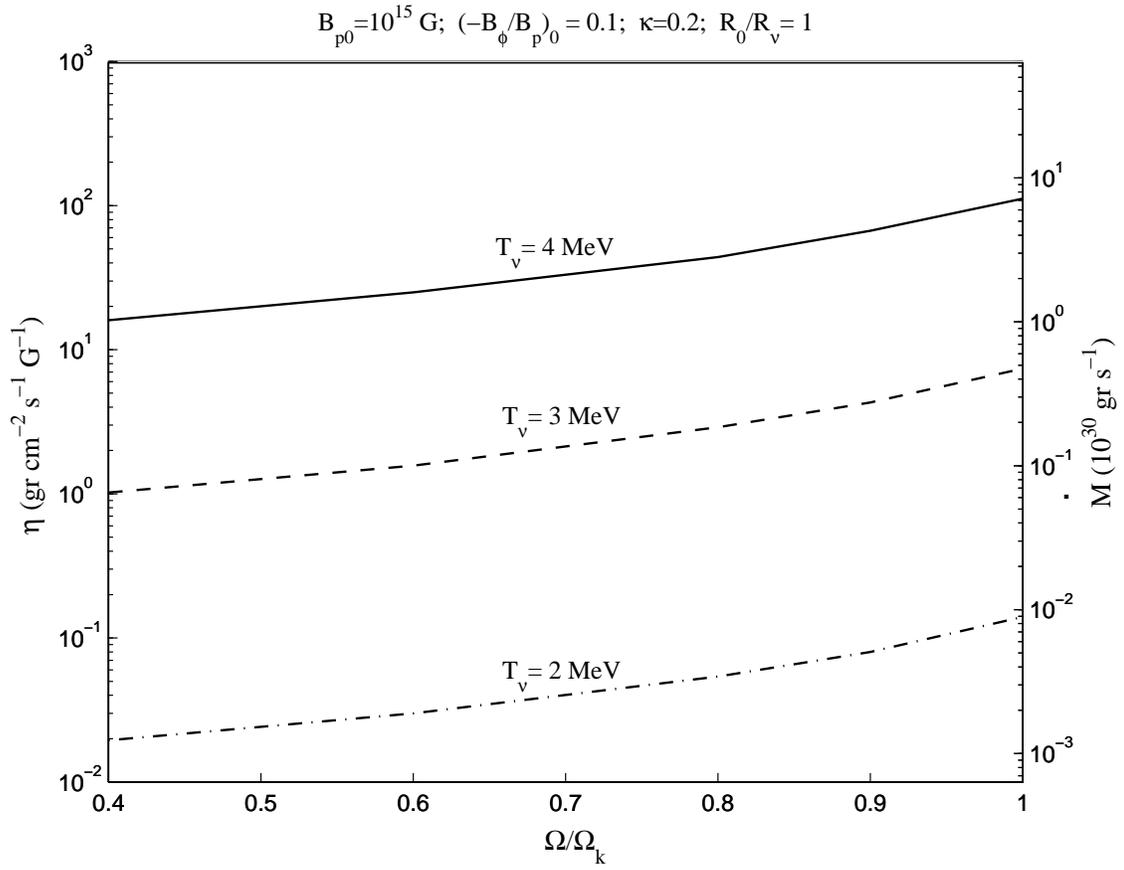}
\caption{$\eta$ and $\dot{M}$ versus angular velocity of the field line, $\Omega$,
given in units of the Keplerian angular velocity at the disk midplane $\Omega_k$.}
\label{f7}
\end{figure}

\clearpage
\begin{figure}[f8]
\plotone{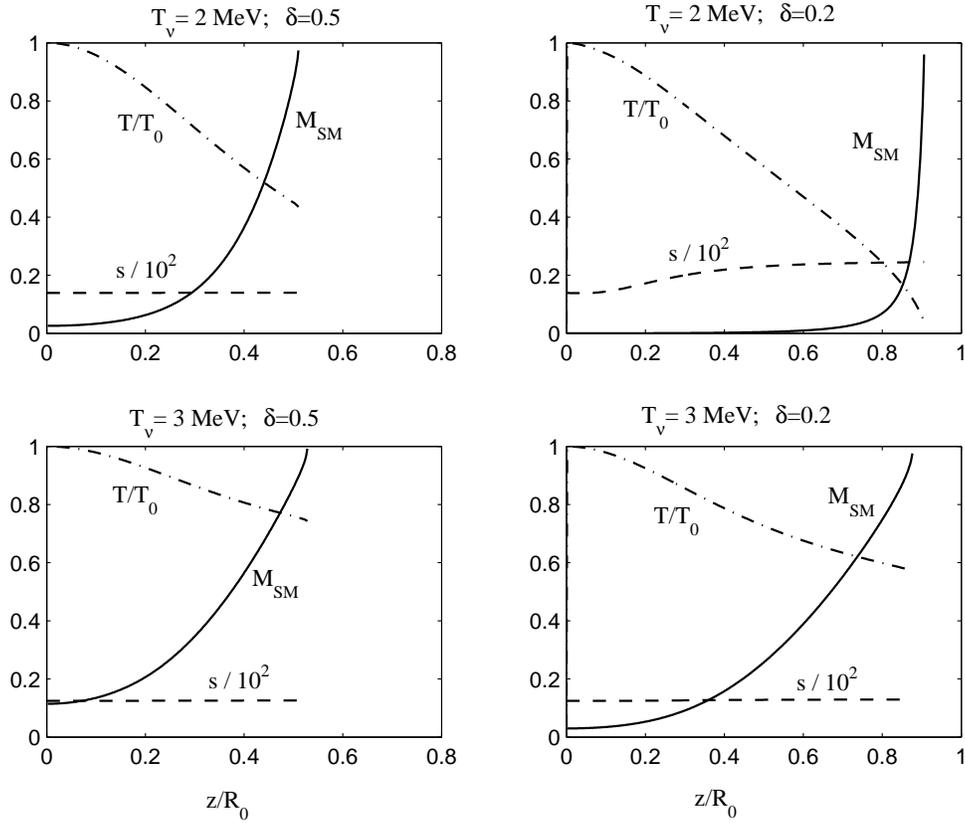}
\caption{Same as fig. 2, but for a self-similar geometry with the parabolic 
field lines given in eq. (\ref{parabolic}).}
\label{f8}
\end{figure}

\end{document}